\documentclass[vecarrow, graybox, natbib]{svmult}

\usepackage{amsmath}
\usepackage{mathptmx}       % selects Times Roman as basic font
\usepackage{helvet}         % selects Helvetica as sans-serif font
\usepackage{courier}        % selects Courier as typewriter font
\usepackage{type1cm}        % activate if the above 3 fonts are
\usepackage{makeidx}         % allows index generation
\usepackage{graphicx}        % standard LaTeX graphics tool
\usepackage{multicol}        % used for the two-column index
\usepackage[bottom]{footmisc}% places footnotes at page bottom

\usepackage{verbatim}
\makeindex             % used for the subject index
\begin{document}

\title*{Modelling Light and Velocity Curves of Exoplanet Hosts}
\author{Rodrigo F.~D\'iaz}
\institute{Rodrigo F.~D\'iaz \at D\'epartement d'Astronomie, Universit\'e de Gen\`eve, Versoix, Switzerland,\\
\email{rodrigo.diaz@unige.ch}\\ \\
Universidad de Buenos Aires, Facultad de Ciencias Exactas y Naturales, Buenos Aires, Argentina\\ \\
CONICET-Universidad de Buenos Aires, Instituto de Astronom\'ia y F\'isica del Espacio (IAFE), Buenos Aires, Argentina
}
\maketitle

\abstract{Research in extrasolar-planet science is data-driven. With the advent of radial-velocity instruments like HARPS and HARPS-N, and transit space missions like \textit{Kepler}, our ability to discover and characterise extrasolar planets is no longer limited by instrumental precision but by our ability to model the data accurately.
This chapter presents the models that describe radial-velocity measurements and transit light curves. I begin by deriving the solution of the two-body problem and from there, the equations describing the radial velocity of a planet-host star and the distance between star and planet centres, necessary to model transit light curves. Stochastic models are then presented and I delineate how they are used to model complex physical phenomena affecting the exoplanet data sets, such as stellar activity. Finally, I give a brief overview of the processes of Bayesian inference, focussing on the construction of likelihood functions and prior probability distributions. In particular, I describe different methods to specify ignorance priors.}

\section{Introduction}
\label{sect.intro}

In the scientist dialogue with Nature data play a crucial role. It is the language by which we receive information from our experiments and observations. Nature communicates with scientists exclusively through data. Without data, there is no science. To be able to interpret the messages encoded in the language of physical observations, we use mathematical models. It is probably useless to stare at a series of data points or samples from a distribution without an underlying idea, however vague, of what the mechanism producing them could be. This idea can be flexible and allow for modifications, but it is an unavoidable step in translating data to knowledge. Roughly speaking, the idea of how data are produced by Nature is a model. More precisely, when the model is expressed using mathematical concepts and terminology it becomes a mathematical model.

Mathematical models consist of one or more equations relating independent to dependent variables, for example time and the radial velocity of a star at that time. Both of these can be measured in an experiment or by an astronomical observation. These equations have constants that represent properties of the system being described, and are usually referred to as the model \emph{parameters}. The basic task of data analysis is to obtain information on model parameters from a set of observations or measurements.

Here we are interested in studying mathematical models of extrasolar-planet data. In particular, we will focus on radial-velocity data and transit light curves. To fully describe these data sets, one needs not only to consider the physics describing the movement of the planets and stars, as well as the geometry of the occultation of a fraction of the stellar disc by the planet, but also the mechanisms producing the actual observed data and their uncertainties, plus all sources of additional radial velocity or photometric variability not of primary interest to us. Actually, at the present stage of exoplanet research, the data are of good enough quality to permit the detection and characterisation of planets of very small mass and size, similar to Earth. However, the signals produced by this type of planetary companions are usually comparable or even smaller than the effects produced by other sources, such as stellar activity, which in addition are usually much more difficult to model. As we will see, the signals we are interested in are relatively simple to model but the determination of the parameters can be hindered or biased by the more complicated effects. Our ability to detect and characterise extrasolar planets is no longer limited by the precision of our instruments but by our ability to model and analyse the data. 

Generally speaking, the mathematical model of datum $d_i$ (which we assume taken at time $t_i$, although this is not fundamental to our formulation) consists of two parts: a prediction of the value the datum $d_i$ should take, $m_i$, and an error term, $e_i$, that accounts for the uncertainties in the data and quantifies the possible discrepancies between the $m_i$ and the datum $d_i$. In general, we can write:
\begin{equation}
\label{eq.model}
d_i = m_i + e_i \, .
\end{equation}
As we will describe in detail in Sect.~\ref{sect.statisticalmodels}, the mathematical formulation leading to $m_i$ can be deterministic or statistical. In the deterministic case, for a certain set of parameters on which the model depends there is no uncertainty in the predicted value $m_i$. The second term, $e_i$, will be assumed to come exclusively from a statistical model of the data. This model can be as simple as the mathematical model corresponding to a series of independent Gaussian variables, but it can be as complicated as one needs to correctly represent the data. This includes models with correlation between the measurements, mixture models etc.~(see Sect.~\ref{sect.gp}).

In this chapter, I will discuss the mathematical models describing the production of two of the most important data types in exoplanet science: radial-velocity measurements and transit light curves (although the focus will be mostly on the radial-velocity models). In Sect.~\ref{sect.models}, I describe the physical models of the orbiting planetary companions, focussing in the case of single-planet systems, that permit an exact analytical description (under certain assumptions). I discuss in Sect.~\ref{sect.statisticalmodels} the stochastic models needed to describe the error term $e_i$ and more complicated phenomena not easily amenable to an analytical description. Finally, Sect.~\ref{sect.bayesian} presents a brief introduction to Bayesian inference and the processes involved in obtaining information on the model parameters from a set of observations. %In Section~\ref{sect.conclusions} I give a summary of the most important points addressed, briefly mention uncovered subjects, and conclude. {\bf [ACA MIRA BIEN SI SECT 5 CONINCIDE]}

\section{Physical models}
\label{sect.models}
In this section, I describe  physical models used to reproduce the radial-velocity time series and light-curve data of a star hosting a planetary companion. In both cases, to reach an analytical expression we need to make a series of assumptions and approximations. Those common to both types of data are:
\begin{enumerate}
\item that the movement of the bodies follows the laws of Newtonian dynamics, and
\item that the bodies do not have electrical charge or any other property besides mass that allows them to interact with each other.
\end{enumerate}
These approximations are usually good in the range of velocities, masses, and distances we will be dealing with.

\subsection{Radial velocities}
\label{sect.rv}
To model radial-velocity data we make one additional assumption: the bodies are dimensionless point particles. This means that the bodies have no internal structure and therefore do not experiment any kind of tidal force. Although this is of course wrong, as we know stars and planets have interiors and processes may occur inside them, the resulting forces are usually much weaker than the gravitational attraction between the bodies and therefore in most cases only produce observable effects over very long time scales \citep[e.g.,][]{zahn77, hut1981}.

Under these assumptions, the movement of the planet and the star reduces to the two-body problem, a traditional problem in classical mechanics. In particular, to obtain a model for the observed radial velocities, we need to obtain the position of the bodies in time. This is known as the Kepler problem. In this section, I follow closely the presentation of the two-body problem by \citet{murraydermott2000} and \citet{murraycorreia2010}. An approach using Lagrangian mechanics is given by \citet{goldstein}.

\subsubsection{Elliptical motion}
\begin{figure}[t]
\begin{center}
%\sidecaption[t]
\includegraphics[scale=0.45]{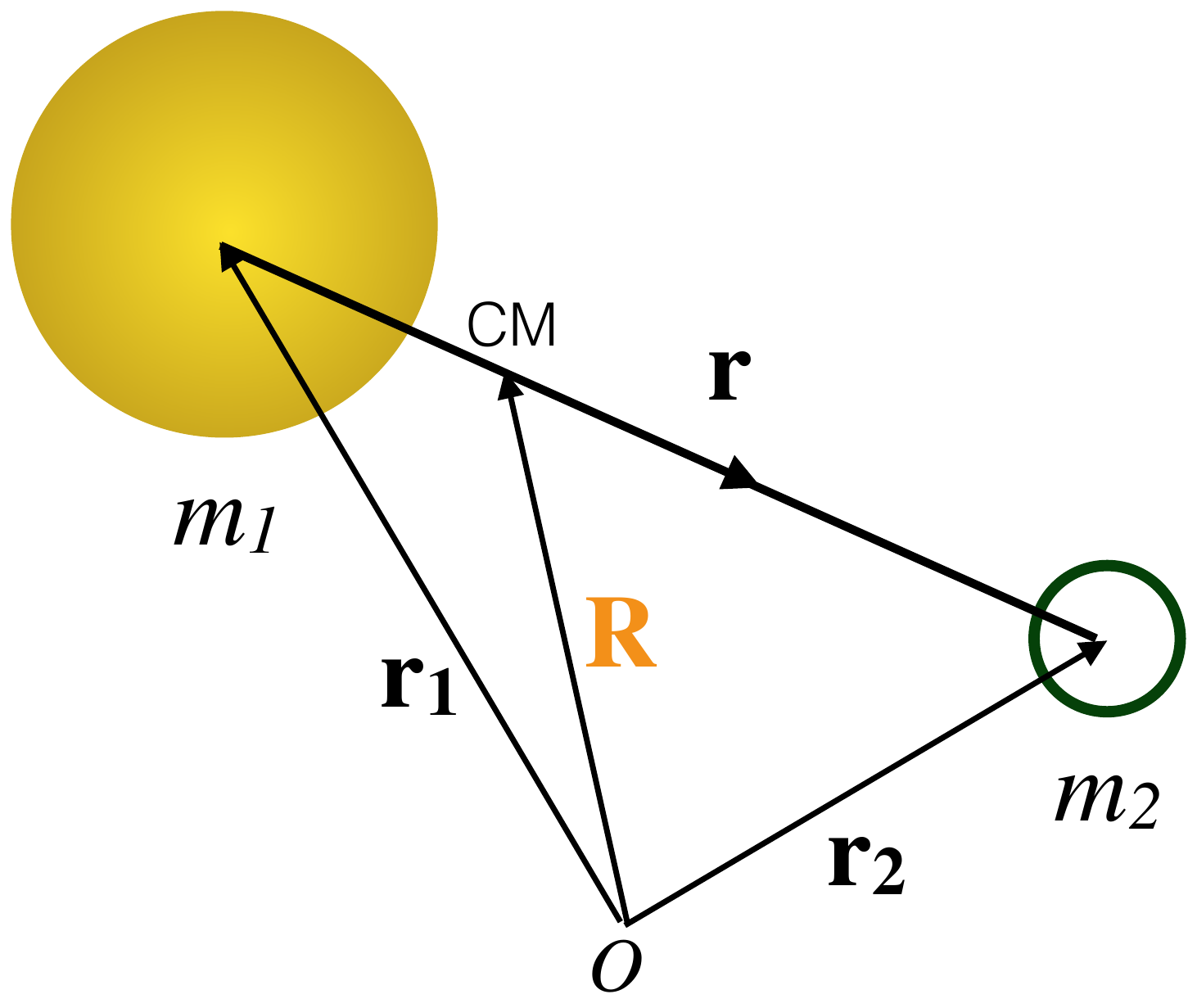}
\caption{Schematic view of the forces and positions in the two-body problem.}
\label{fig.twobody}
\end{center}
\end{figure}

Let us consider two objects of masses $m_1$ (the star) and $m_2$ (the planet) interacting gravitationally. Their equations of motion are described in an inertial frame $S$ with origin at $O$:
\begin{align}
\label{eq.f1}
\vec{F_1} &= m_1 \ddot{\vec{r_1}} = +G \frac{m_1 m_2}{r^3} \vec{r}\,,\\
\label{eq.f2}
\vec{F_2} &= m_2 \ddot{\vec{r_2}} = -G \frac{m_1 m_2}{r^3} \vec{r}\,,
\end{align}
where $G$ is the Universal gravitational constant and $\vec{r} = \vec{r_2} - \vec{r_1}$ is the relative position vector, pointing from the star to the planet (see Fig.~\ref{fig.twobody}).
Dividing Eq.~(\ref{eq.f1}) by $m_1$ and Eq.~(\ref{eq.f2}) by $m_2$, these equations can be combined to produce the equation of relative motion:
\begin{equation}
\ddot{\vec{r}} + G (m_1 + m_2)\frac{\vec{r}}{r^3} = 0\,.
\label{eq.relmot}
\end{equation}

\begin{comment}
The Universal Law of Gravitation enunciated by Newton states that any two objects with masses $m_1$ and $m_2$ separated by a distance $r$ will experience a mutual attractive force whose magnitude is

$$
F = G \frac{m_1 m_2}{r^2}\;\;,
$$
where $G$ is the Universal gravitational constant. %, with units of $\mathrm{s^{-2} m^3/kg}$, i.e. an inverse density over a squared time. These units shed a light on what physical magnitudes will be measurable from gravitational effects (see Sect.~\ref{sect.nonkeplerian}). 
Considering an inertial frame $S$ with origin at $O$ (Fig.~\ref{fig.twobody}), we can write the equations of motion of both bodies in vectorial form:
\begin{align}
\label{eq.f1}
\vec{F_1} &= m_1 \ddot{\vec{r_1}} = +G \frac{m_1 m_2}{r^3} \vec{r}\;\;,\\
\label{eq.f2}
\vec{F_2} &= m_2 \ddot{\vec{r_2}} = -G \frac{m_1 m_2}{r^3} \vec{r}\;\;,
\end{align}
where $\vec{r} = \vec{r_2} - \vec{r_1}$ is the relative position vector, pointing from the star to the planet (Fig.~\ref{fig.twobody}).
Dividing Eq.~\ref{eq.f1} by $m_1$ and Eq.~\ref{eq.f2} by $m_2$ equations can be combined to produce the equation of relative motion:
\begin{equation}
\ddot{\vec{r}} + G (m_1 + m_2)\frac{\vec{r}}{r^3} = 0\;\;.
\label{eq.relmot}
\end{equation}

\end{comment}

Taking the vector product of Eq.~(\ref{eq.relmot}) with the relative position vector $\vec{r}$, and using the fact that $\vec{r} \times \vec{r} = 0$, we find the first conserved magnitude of the two-body problem: $\vec{r}\times\ddot{\vec{r}} = 0$,
which is promptly integrated to give
\begin{equation}
\vec{r}\times\dot{\vec{r}} = \vec{h}\,.
\end{equation}
The vector $\vec{h}$ is constant and, because it is the vector product of $\vec{r}$ and $\dot{\vec{r}}$, it is perpendicular both to the position and velocity vectors. This means that the motion of the system happens in a plane perpendicular to the constant vector $\vec{h}$, known as the orbital plane. It is then practical to express the position and velocity vector in polar coordinates on this plane: $\vec{r} = r \hat{r}$, $\dot{\vec{r}} = \dot{r}\hat{r} + r\dot{\theta}\hat{\theta}$, where $\hat{r}$ is the unit vector pointing from the star to the planet and $\hat{\theta}$ is a unit vector perpendicular to $\hat{r}$. We then find that
\begin{equation}
\vec{h} = r^2 \dot{\theta}\hat{z}\,,
\end{equation}
where $\hat{z}$ is a unit vector forming a right-handed triad with $\hat{r}$ and $\hat{\theta}$. This is a good approximation of angular momentum of the system when\footnote{As we are describing the relative motions of the bodies, the \emph{actual} angular momentum of the system contains a term 	related to the motion of the star. Whenever $m_2\ll m_1$, this term can be neglected and $h$ is equal to the total angular momentum of the system per unit mass of the body $m_2$.} $m_2 \ll m_1$. From the conservation of $\vec{h}$, one can deduce the second law of planetary motion discovered by Kepler \citep[see][sect.~2.2]{murraydermott2000}, which states that equal areas are swept out by the position vector in equal times.  Additionally, it can be shown that the area swept by unit time is 
\begin{equation}
\frac{\mathrm{d}A}{\mathrm{d} t} = \frac{h}{2}\,.
\end{equation}

\begin{figure}[t]
\begin{center}
%\sidecaption[c]
\includegraphics[scale=0.7]{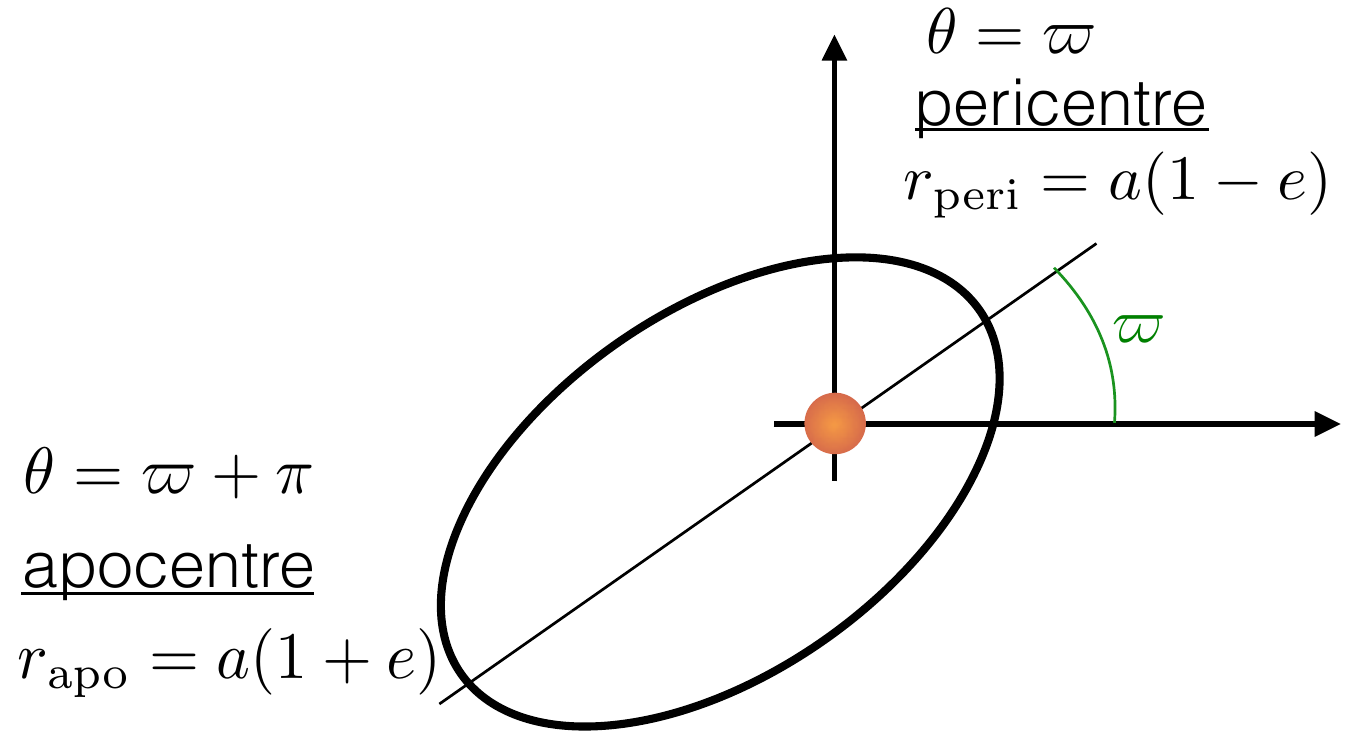}
\caption{Schematic view of the trajectory of an orbiting companion to the central star (at the origin of coordinates), as seen vertically from above the orbital plane. The pericentre, apocentre, and their corresponding distances are indicated.}
\label{fig.ellipse}
\end{center}
\end{figure}

With this conserved quantity, we can express the general solution of Eq.~(\ref{eq.relmot}):
\begin{equation}
\label{eq.conic}
r = \frac{p}{1 + e \cos(\theta - \varpi)}\,,
\end{equation}
where
\begin{equation}
p = \frac{h^2}{G(m_1 + m_2)}
\label{eq.p}
\end{equation}
and $e$ and $\varpi$ are two constants of integration. Equation (\ref{eq.conic}) is the general equation of a conic section in polar coordinates. In particular, whenever $0<e<1$ and $p = a(1-e^2)$, the equation describes an ellipse of eccentricity $e$ and semi-major axis $a$:
$$
r = \frac{a (1 - e^2)}{1 + e \cos(\theta - \varpi)} \, .
$$
 This is one possible solution for the movement of a planet around a star. The two extreme distances, called periapsis and apoapsis occur when the $\theta = \varpi$ and $\theta = \pi + \varpi$, respectively, and are equal to $a(1-e)$ and $a(1+e)$ (see Fig.~\ref{fig.ellipse}). We can define a new polar angle, $\nu$, such that $\nu = \theta - \varpi$, i.e., we measure the angles starting at the periapsis. This angle usually receives the name of true anomaly. Then,
\begin{equation}
\label{eq.ellipse}
r = \frac{a (1 - e^2)}{1 + e \cos\nu}\,.
 \end{equation} 

The area enclosed by the ellipse, $\pi a b$, where $b$ is the semi-minor axis of the ellipse, $b = a(1-e^2)$, is swept by the position vector in one orbital period, $P$. From Kepler's second law, and the fact that $h^2 = G(m_1+m_2)a(1-e^2)$ (cf.~Eq.~\ref{eq.p}), we arrive at Kepler's third law of planetary motion:
\begin{equation}
P^2 = \frac{4\pi^2}{G(m_1+m_2)}a^3\,.
\end{equation}

\subsubsection{Barycentric orbits}
Combining Eqs.~\ref{eq.f1} and \ref{eq.f2}, one obtains the equation of motion for the centre of mass (CM) of the system:
\begin{equation}
m_1 \ddot{\vec{r_1}} + m_2 \ddot{\vec{r_2}} = 0\,,
\end{equation}
which can be easily integrated to obtain the equation describing the position of the CM in time:
\begin{equation}
\vec{R}(t) = \frac{\alpha t + \beta}{m_1 + m_2}\,,
\end{equation}
where $\vec{R} = (m_1 \vec{r_1} + m_2 \vec{r_2})/(m_1 + m_2)$ is the position of the CM of the system (Fig.~\ref{fig.twobody}). As expected from the conservation of momentum, the CM of the system moves with a constant velocity with respect to the origin $O$. Since the original frame of reference $S$ was assumed inertial, this means that a reference frame with origin in the CM is also inertial. We will find it useful to describe the motions of the planet and the star in this reference frame. We can then write the positions of the masses with respect to the CM as
\begin{comment}
$$
\vec{R_1} = \vec{r_1} - \vec{R}\;\;, \qquad \vec{R_2} = \vec{r_2} - \vec{R}\;\;,
$$
and also write the positions of the masses in frame $S$ as
$$
\vec{r_1} = \vec{R} - \frac{m_2}{m_1 + m_2}\vec{r}\;\;,\qquad
\vec{r_2} = \vec{R} + \frac{m_1}{m_1 + m_2}\vec{r}\;\;,
$$
from which we conclude that
\end{comment}
\begin{align}
\label{eq.bigr1}
\vec{R_1} &= - \frac{m_2}{m_1 + m_2}\vec{r}\;\;,\\
\label{eq.bigr2}
\vec{R_2} &= + \frac{m_1}{m_1 + m_2}\vec{r}\;\;.
\end{align}
This means that $\vec{R_1}$ and $\vec{R_2}$ always have opposite directions and that the CM is in the line joining the two vectors. Besides, it implies that the trajectories of the star and the planet with respect to the barycentre of the system are simply scaled down versions of the conic section describing the relative motion of the bodies (Eq.~\ref{eq.ellipse}). In the case of the ellipse, the semi-major axis of the star and planet orbital trajectories are scaled down by a factor $m_2/(m_1+m_2)$ and $m_1/(m_1+m_2)$, respectively. Note that in the typical case where $m_2 \ll m_1$, the stellar orbit with respect to the CM becomes very small while the planet orbit resembles the relative orbit.

\subsubsection{Projection and radial velocity}
\label{sect.projection}
The orientation of the orbit in three-dimensional space can be described by three angles. The angle of the orbital plane with respect to the plane of the sky is the orbital inclination $I$. The intersection between the plane of the sky and the orbital plane is called the line of nodes. The longitude of the ascending node, $\Omega$, is the angle between a reference direction in the plane of the sky and the radius vector at the ascending node, where the planet crosses the plane of the sky from below to above. Finally, the argument of periapsis, $\omega$, is the angle between that same radius vector and the orbital periapsis measured on the orbital plane. For $I \sim 0$, we have $\varpi = \omega + \Omega$, where $\varpi$ is the two-body constant already found in Eq.~(\ref{eq.conic}).

To obtain the expected radial velocity of the star in the system, let us project the expression of the orbit to a cartesian coordinate system centred in the CM of the system and oriented so that the positive $z$ axis points towards an observer on Earth and the $xy$ plane corresponds to the plane of the sky. To simplify the resulting expressions, we further assume that the reference direction defined by the positive $x$ axis coincides with the line of nodes (i.e., $\Omega = 0$). Under these conditions, the cartesian components of the barycentric movement of the stellar body are:
\begin{align}
\label{eq.x}
x &= r_1 \cos(\omega + \nu)\,,\\
\label{eq.y}
y &= r_1 \sin(\omega + \nu)\cos I\,,\\
\label{eq.z}
z &= r_1\sin(\omega + \nu)\sin I\,,
\end{align}
with $$r_1 = a \frac{m_2}{m_1+m_2} \frac{(1-e^2)}{1 + e \cos\nu}$$ the distance between the star and the CM of the system. Deriving Eq.~(\ref{eq.z}) with respect to time and using the fact that 
$$
\dot{\nu} = \dot{\theta} = h/r^2 = \frac{2\pi}{P}\frac{a^2\sqrt{1 - e^2}}{r^2}\,,
$$
we get the expression of the stellar radial velocity $V$ as a function of the true anomaly:
\begin{equation}
\label{eq.rv}
V = V_0 + K\left[\cos\left(\nu+\omega\right) + e\cos\omega\right]\,,
\end{equation}
with
\begin{equation}
\label{eq.amplitude}
K = \left(\frac{2\pi G}{P}\right)^{1/3}\frac{1}{\sqrt{1-e^2}}\frac{m_2 \sin I}{\left(m_1+m_2\right)^{2/3}}\,,
\end{equation}
and $V_0$ and integration constant corresponding to the velocity of the CM with respect to the observer on Earth.

If we neglect $m_2$ with respect to $m_1$ in the denominator of Eq.~(\ref{eq.amplitude}), which is reasonable in the typical case $m_2 \ll m_1$, we see that the radial-velocity signal scales linearly with the mass $m_2$ of the orbiting companion, and inversely with $P^{1/3}$. These two facts introduce strong biases in the sample of detected planets and in the shapes of the distributions of their orbital parameters. Note that the RV amplitude depends on the combination of $m_2$ and $\sin I$. Using RV data alone it is therefore not possible to measure the real mass of the planets, but only the so-called ``minimum mass'', $m_2\sin I$.

\subsubsection{Kepler problem}
Up to this point we have obtained a description of the trajectories followed by the bodies of a star-planet system and have reached an expression for the radial velocity of the star as a function of the position of the planet in the orbit, $\nu$. What remains to be done to link this expression (Eq.~\ref{eq.rv}) with the observations is to describe the motion of the bodies in time as they traverse their orbits. In other words, we need to find the function $\nu = \nu(t)$. This is known as the Kepler problem and is much more involved than obtaining the equation of the orbit or the radial-velocity expression. We give here a summarised description of the solution to the Kepler problem and refer the interested reader to specialised literature \citep[e.g.,][]{goldstein,murraydermott2000}.

The integrals involved in finding $\nu(t)$ are more easily solved using two auxiliary variables: the eccentric anomaly, $\psi$, defined through the expression
\begin{equation}
\label{eq.ea}
r = a(1 - e\cos\psi)\,,
\end{equation}
and the mean anomaly, $\mu$,
$$
\mu = \frac{2\pi}{P}(t - \tau)\,,
$$
where $\tau$ is the time of passage through the periapsis\footnote{Of course, exactly as we have redefined the polar angle so that $\nu$ is 0 at periapsis, we could also measure time starting at the moment of periastron passage, and get rid of $\tau$ in the definition of the mean anomaly. However, usually $\tau$ is unknown and including it as a model parameter allows us to measure it.}. Note that the time $t$ enters explicitly in the definition of $\mu$. 

It can be shown that $\mu$ and $\psi$ are related through the Kepler equation:
\begin{equation}
\mu = \psi - e\sin\psi\,,
\end{equation}
which is transcendental and requires iterative methods to solve it\footnote{A \texttt{Python} code to compute the true anomaly from the mean anomaly and eccentricity has been made available at \url{https://github.com/exord/faial/blob/master/trueanomaly.py}.}. By inverting the Kepler equation, one can find $\psi$ and the radial distance $r$ as a function of time (Eq.~\ref{eq.ea}). Furthermore, comparing the defining equation (Eq.~\ref{eq.ea}) with the equation of the elliptical orbit (Eq.~\ref{eq.ellipse}) one can find, after some algebra, an expression for $\nu$ as a function of $\psi$:
\begin{equation}
\tan\left(\frac{\nu}{2}\right)= \sqrt{\frac{1+e}{1-e}}\,\tan\left(\frac{\psi}{2}\right)\,.
\end{equation}

We have shown how to obtain the time dependence of the true anomaly $\nu$, and therefore of the radial-velocity expression (Eq.~\ref{eq.rv}), which can be promptly used to compare the model with the observations (see Sect.~\ref{sect.bayesian}).

\subsection{Transits}
When the orbital inclination $I$ is close to 90 degrees, the orbiting planet, as seen from Earth, passes in front of the stellar disk, causing a slight dimming of the star known as a transit. This event is extremely rich in information about the star-planet system. In particular, it allows lifting the $m_2\sin I$ degeneracy and measure the real planet mass.

Under certain assumptions, the model of the light curve of a transiting planet can be written analytically. We need to abandon the assumption of point masses used for the radial-velocity model, but assume both the planet and the star are perfectly spherical with radii $R_{\rm p}$ and $R_{\rm s}$, respectively. Additionally, we assume that the planet is completely opaque and emits no radiation.

\subsubsection{Uniform source}
\newcommand{\df}{\sqrt{\Delta F}}
\newcommand{\tfull}{\omega_F}
\newcommand{\ttot}{\omega_T}
\newcommand{\sinratio}{\sin^2\tfull/\sin^2\ttot}
\newcommand{\deltaplus}{\left(1 + \df\right)^2}
\newcommand{\deltaminus}{\left(1 - \df\right)^2}
\newcommand{\dist}{\delta}

\begin{figure}[t]
\includegraphics[scale=1]{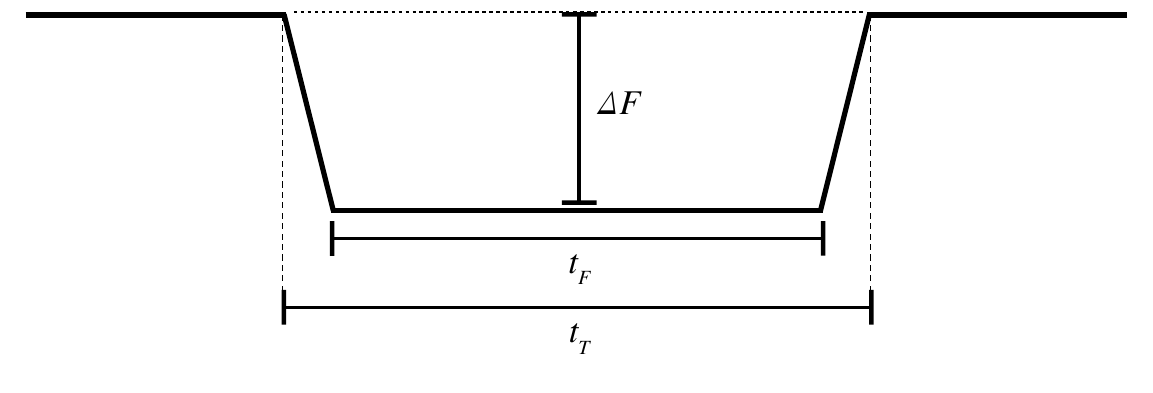}
\caption{Schematic view of a flat-bottomed transit. Three of the four parameters describing the model are indicated: the transit depth $\Delta F$, the total transit duration, $t_T$, and the duration of the flat part, $t_F$. The remaining parameter is the orbital period $P$.}
\label{fig.flatbottom}
\end{figure}

Further assuming that the star is a uniform source, we can write the expression of the ratio between obscured to unobscured flux, $F(k, \dist) = 1 - \lambda(k, \dist)$, where $\lambda$ is the flux loss as a function of the radius ratio $k=R_{\rm p}/R_{\rm s}$ and $\dist = d/R_{\rm s}$ is the projected centre-to-centre distance between the star and the planet, normalised by the radius of the star. Following \citet{mandelagol2002}, and assuming $k < 1$, we find:
%\begin{align*}
\begin{equation}
\label{eq.uniform}
\lambda(k, \dist) = 
\begin{cases}
0, & \dist>1+k\,,\\
\frac{1}{\pi}\left[k^2\kappa_0 + \kappa_1 - \sqrt{\frac{4\dist^2 - \left(1 + \dist^2 - k^2\right)^2}{4}}\right], &1 - k < \dist \leq 1+k\,,\\
k^2, &\dist\leq 1-k\,,\\
\end{cases}
\end{equation}
%\end{align*}
where 
\begin{equation}
\kappa_0 = \cos^{-1}\left[\frac{k^2 + \dist^2 -1}{2k\dist}\right]\;\;,\qquad \kappa_1 = \cos^{-1}\left[\frac{1 - k^2 + \dist^2}{2\dist}\right]\,.
\end{equation}
In other words, there is no flux loss as long as $d > R_{\rm p} + R_{\rm s}$, and the loss is constant and equal to $(R_{\rm p}/R_{\rm s})^2$ when the planet disc is completely inside the stellar disc (see Fig.~\ref{fig.flatbottom}).

If the planet is on a circular orbit, \citet{seagermallen-ornelas2003} showed there is a unique relation between the four parameters describing the model (see Fig.~\ref{fig.flatbottom}) and some combinations of the physical parameters of the system. This allows obtaining, in this simple case, information about the planetary system by measuring the transit properties:
\begin{align}
\label{eq.radiusratio}
\frac{R_{\rm p}}{R_{\rm s}} &=\df\,,\\
\label{eq.impact}
b = \frac{a}{R_{\rm s}}\cos I &= \left[\frac{\sin^2{\ttot}\deltaminus - \sin^2{\tfull}\deltaplus}{\sin^2{\ttot} - \sin^2{\tfull}}\right]^{1/2}\,,\\
\frac{a}{R_{\rm s}} &=\left[\frac{\left(1 + \df\right)^2 - b^2\left(1 - \sin^2{\ttot}\right)}{\sin^2{\ttot}}\right]^{1/2}\,,
\label{eq.ars}
\end{align}
where $\tfull = t_F \pi / P$ and $\ttot = t_T \pi / P$. Note that, combining Eqs.~(\ref{eq.impact}) and (\ref{eq.ars}), one can measure the orbital inclination $I$. Combining this information with RV data providing $m_2 \sin I$ one can measure the true planet mass $m_2$. Interestingly, using Kepler's third law, we can write an expression involving the mean densities of the planet and the star:
\begin{equation}
\rho_{\rm s} + k^3 \rho_{\rm p} = \frac{3\pi}{G}\frac{1}{P^2}\left(\frac{a}{R_{\rm s}}\right)^3\,,
\label{eq.stellardens}
\end{equation}
where the mean density is $\rho_{\rm s} = M_{\rm s} \left(4 \pi R_{\rm s}^3/3\right)^{-1}$, and an equivalent expression for $\rho_{\rm p}$. In the usual case in which $k \ll 1$ and $M_{\rm p} \ll M_{\rm s}$, the mean density of the star can be approximately obtained using Eq.~(\ref{eq.ars}):
\begin{equation}
\rho_{\rm s} \approx \frac{3\pi}{GP^2}\left[\frac{\left(1 + \df\right)^2 - b^2\left(1 - \sin^2{\ttot}\right)}{\sin^2{\ttot}}\right]^{3/2}\,, 
\label{eq.stellardens2}
\end{equation}
valid only in the case of circular orbits.

The transit duration is usually measured in hours while the orbital period $P$ is usually of the order of days. Therefore, we see that the parameters $\ttot$ and $\tfull$ are necessarily small. In this case the sine function can be approximated $\sin^2{x} \approx x^2$, and Eqs.~(\ref{eq.impact}) and (\ref{eq.ars}) can be simplified:
\begin{align}
b &\approx \left[\frac{t_T^2\deltaminus - t_F^2\deltaplus}{t_T^2 - t_F^2}\right]^{1/2}\,,\\
\frac{a}{R_{\rm s}} &\approx \frac{2P}{\pi} \frac{\Delta F^{1/4}}{\sqrt{t_T^2 - t_F^2}}\,.
\end{align}

In the case of non-circular orbits, complicated algebra arises and it is no longer possible to find exact analytical expressions for the physical system parameters. \citet{winn2008} gives approximate expressions valid when $R_{\rm p}\ll R_{\rm s}\ll a$. The expression for the radius ratio (Eq.~\ref{eq.radiusratio}) and for the impact parameter (Eq.~\ref{eq.impact}) remain the same, but the normalised semi-major axis is scaled by a factor $\sqrt{1 - e^2} / \left(1 + e\sin{\omega}\right)$:
\begin{equation}
\frac{a}{R_{\rm s}} \approx \frac{2P}{\pi} \frac{\Delta F^{1/4}}{\sqrt{t_T^2 - t_F^2}}\left(\frac{\sqrt{1 - e^2}}{1 + e\sin{\omega}}\right)\,.
\end{equation}
Depending on the orientation of the orbit in the sky, the semi-major axis can be smaller or larger than the one measured in the circular case. As a consequence, the mean stellar density computed by means of Eq.~(\ref{eq.stellardens2}) can be under- or overestimated if the orbit is incorrectly assumed circular. Further analysis in the case of eccentric orbits is presented by \citet{kipping2008}.

\subsubsection{Non-uniform source}
In reality, stars are not observed as uniform disks of light. Instead, their brightness appears to decrease from the centre to the limb of the disc, an effect known as limb darkening. \citet{mandelagol2002} provide analytical equations of the flux drop, $F$, as a function of the normalised star-to-planet centre separation, $\dist$, for a quadratic and non-linear limb-darkening laws, which most accurately describe the observed darkening effect \citep{claret2000}. Their codes are available to the community and are widely used by researchers in the field.

When limb darkening is included in the transit model, the physical parameters become strongly covariate and the model is degenerate, which complicates the process of inferring the system parameters from data (see Sect.~\ref{sect.bayesian}).

\subsubsection{Projected distance as a function of time}
The model of the drop in flux produced by the transit of a planetary object is given as a function of the projected centre-to-centre normalised distance, $\dist = d/R_{\rm s}$, where $d$ is the sky-projected distance between the centres of the planet and the star. To obtain a model that we can compare directly\footnote{Special care must be taken when the timescale of the variability is comparable to the integration time of individual points \citep{kipping2010}.} to a time series of flux measurements, we need to express $\dist$ as a function of time.

This is promptly achieved by considering the projection of the relative orbit into cartesian coordinates described in Sect.~\ref{sect.projection}. Recalling that we had chosen the axes so that the $xy$ plane coincided with the plane of the sky, we can write:
\begin{equation}
\delta = \frac{1}{R_{\rm s}}\left(x^2 + y^2\right) = \frac{a}{R_{\rm s}}\frac{1-e^2}{1+e \cos{\nu}}\sqrt{1 - \sin^2{\left(\omega + \nu\right)}\sin^2{I}}\,,
\end{equation}
where we have replaced the distance $r_2$ by the relative distance $r$ ($a$ is the semi-major axis of the relative orbit). Substituting this expression in the models for the drop of flux, and using the dependence of the true anomaly $\nu$ with time we can compute a model  transit light curve to compare with the data time series.
 
\subsection{Non-Keplerian models}
\label{sect.nonkeplerian}
Up to here, we have assumed the modelled planetary companion moves in a Keplerian orbit around its host star. This is a good approximation when tidal forces originating in the interior of the objects can be neglected, which is often the case, as discussed above. However, the assumption of Keplerian motion is invalidated by the presence of additional objects in the system. Indeed, if the star is orbited by multiple planets, their mutual gravitational interactions will make their motion depart from the Keplerian orbit. The effect of the planet-planet interactions is usually very small on the radial-velocity data. It requires measurements with extreme precision spanning a long time for these effects to be detectable \citep[see][]{correia2010}. If detected, planet-planet interactions can lift the degeneracy between the RV-measured minimum masses and the orbital inclination $I$.

Gravitational interactions between the planets are more easily detected in transit light curves through the measurement of the timing of the transits, and their departure from perfect periodicity. This effect, known as transit timing variations (TTV) was predicted and studied for years \citep[e.g.,][]{agol2005, holmanmurray2005} before the detection of the first transit system with clear TTV \citep{holman2010}. Besides, planet-planet interactions can change the duration of the transits, as the planets cross the stellar discs at different latitudes at different times. This gives rise to transit duration variations or TDV. Timing methods have  been used to measure the mass of transiting planets for which no radial velocity measurement was possible \citep[e.g.,][]{jontof-hutter2015} and to accurately measure the physical parameters of systems independently of stellar models \citep{almenara2015, almenara2016}. TTV of a transiting planet can also be employed to detect non-transiting planets \citep[e.g.,][]{ballard2011}. However, unless TDV are also present \citep[see][]{nesvorny2013, barros2014}, the parameters of the unseen planet are usually degenerate \citep{ballard2011}.

Most of the analysis on TTV and TDV rely on the measurement of timings and durations of individual transits, and are limited by the precision obtained in these measurements. A way to dramatically improve the precision obtained in the parameters of systems analysed through transit timing is to perform a full photodynamical model \citep[e.g.,][]{carter2011, doyle2011, almenara2015, almenara2016}. The full photodynamical model uses a dynamical $N$-body simulation to compute distances between all the planets and the star at each time step. This is then input in the light curve models discussed above. In this way, the timing of each individual transit is constrained by the data from the entire light curve and not just by the points taken during that particular transit. As a consequence, analyses using photodynamical models produce superior results than those using measurements from individual transits \citep{almenara2016}, but at the expense of the increased computational time required to perform the dynamical simulation.

\section{Statistical models}
\label{sect.statisticalmodels}
Let us go back to Eq.~(\ref{eq.model}):
\begin{equation}
d_i = m_i + e_i\,.
\label{eq.model2}
\end{equation}
Last section was dedicated to obtaining deterministic physical models for the term $m_i$. To do this we made a series of simplifying assumptions to render the 
problem tractable and arrived at deterministic expressions for the radial velocity or light curve of a star with an orbiting planetary companion. In other words, for each set of model parameters at a given time, the models from the previous sections provide a precise value for the radial velocity or flux drop.

In this section we describe a second type of models used to describe (exoplanet) data: statistical models. In a statistical (or stochastic) model there is not a single model output value for a given set of input parameters and independent variables (time), but rather a probability distribution of values. Stochastic models are strictly necessary when the system being modelled is probabilistic in nature. This is of course the case of quantum mechanics, but also of all problems involving uncertain measurements. 

In Eq.~(\ref{eq.model2}), the error term $e_i$ represents the discrepancy between the model $m_i$ and the observed data. For data with non-zero uncertainty (real data) this term exhibits a stochastic behaviour. Its probabilistic nature comes from a combination of actual probabilistic processes --- such as the emission of radiation by the atoms in the stellar atmosphere --- with very complex, but in principle deterministic, systems --- such as the detailed behaviour of the telescope and instruments used to acquire the data or weather conditions at the time of observation. These complex systems are better described in probabilistic terms instead of using complicated deterministic models with a large number of parameters.

Besides the probabilistic nature of the error term, $e_i$, a second source of randomness comes from uncertainties in the physical model term $m_i$. We have already mentioned that $m_i$ is not necessarily deterministic. Indeed, $m_i$ may contain a probabilistic part related either to uncertainty in the model itself or in the independent variable used to compute it. The latter is discussed in some detail in \citet[][sect.~4.8.2]{gregory}. Here, we will deal exclusively with the former, and will break down the model term as
\begin{equation}
m_i = m^{\rm d}_i + m^{\rm s}_i\,,
\end{equation}
where the $d$ and $s$ indices stand for {\it deterministic} and {\it stochastic}, respectively.

In general, the term $m^{\rm s}_i$ can include complicated processes affecting the data and on which we are not primarily interested. A prime example of this is stellar activity. As we will describe in this section, it is possible to produce a relatively simple model of the effects of stellar activity using statistical models. As we are not primarily interested in studying stellar activity, the parameters of this part of the model can be marginalised out to fully account for their effect on the uncertainty of the remaining model parameters. A deterministic alternative would require producing a model that provides the precise effect of stellar activity on the data for a given time. This usually requires a large number of parameters to describe, for example, the position, size, temperature and spectrum of active regions in time. Although this type of models exist and have been successfully put to practice \citep{boisse2012, kipping2012b, dumusque2014}, they are plagued with degeneracies, because the available data are usually not sufficient to determine all model parameters independently. Besides, these models tend to require a substantial amount of computing time, which renders them impractical for iterative algorithms such as Markov chain Monte Carlo (MCMC). Another typical case where statistical models provide a simpler description than their deterministic counterparts is instrument systematics. In summary, the term $m^{\rm s}_i$ includes all ``complex physics'', defined here as all processes or systems too complicated to be described by a deterministic model with a reasonable number of parameters.

\subsection{Uncorrelated errors}
The simplest possible model for the error term $e_i$ arises when the observed data points are assumed uncorrelated and their errors are normally distributed with known variance $\sigma_i^2$, not necessarily the same for all measurements. The expression for the distribution function of the error term, $f(e_i)$ is then simply:
\begin{equation}
f_{\sigma_i}(e_i) = N(0, \sigma_i) = \frac{1}{\sqrt{2\pi\sigma_i^2}}\exp\left[-\frac{e_i^2}{2\sigma_i^2}\right]\,,
\label{eq.normalerror}
\end{equation}
where $N(\mu, \sigma)$ is a Gaussian distribution with mean value $\mu$ and variance $\sigma^2$.

A slightly more sophisticated model consists in assuming the errors are over- or underestimated, and therefore introducing a multiplicative correction factor $\alpha$:
\begin{equation}
f_{\sigma_i, \alpha}(e_i)= N(0, \alpha \sigma_i) = \frac{1}{\sqrt{2\pi}\alpha\sigma_i}\exp\left[-\frac{e_i^2}{2\alpha^2\sigma_i^2}\right]\,.
\label{eq.normalerror2}
\end{equation}
Note that this model already has an additional nuisance parameter $\alpha$, which in principle is not of primary interest. The $\alpha$ term can be assumed to be the same for all measurements (for all $i$), or depend on subindex $i$, with the subsequent multiplication of nuisance parameters.

\subsection{Gaussian processes}
\label{sect.gp}

A stochastic process can be roughly described as a generalisation of a probability distribution to functions \citep{rasmussenwilliams2005}. In other words, it is the probability distribution of an infinite-dimensional random variable, one that describes the values of the function in all possible points in input space. A Gaussian process (GP) can be seen as a collection of such random variables, any finite number of which have a joint multivariate normal distribution \citep{rasmussenwilliams2005}. Because of their great flexibility, attractive mathematical properties and computational tractability, GPs have been used as models by the machine learning community. Only recently GP regression has made its way to the exoplanet community. A thorough treatment of GPs is outside the scope of this chapter and we therefore give only a brief introduction to the subject in the context of exoplanetary science.

In general, GP regression is used in exoplanetary science to model complicated signals that are difficult to describe analytically. These include, but are not limited to, systematic effects originating in the instruments and signals produced by stellar activity. Sometimes GPs are said to model the covariate noise of a given dataset. This is equivalent, but means the GP is included in the error term $e_i$ instead of in the stochastic model term $m_i^{\rm s}$.

A GP is completely specified by its mean function and covariance function, which we denote $m(\vec{x})$ and $k(\vec{x}, \vec{x^\prime})$, respectively, where $\vec{x}$ and $\vec{x^\prime}$ are input vectors of the process. An input vector can have any number of dimensions, as the process can have any number of input variables. For example, when modelling stellar activity signals, one would naturally include time as an input variable, but may also find it useful to include activity proxies, such as the Ca II H \& K lines index, $\log R^\prime_\mathrm{HK}$. For any real process $f(\vec{x})$, the mean and covariance functions are defined as 
$$
m(\vec{x}) = \mathrm{E}[f(\vec{x})]\;\;\text{and} \quad
k(\vec{x}, \vec{x^\prime}) = \mathrm{E}[(f(\vec{x}) - m(\vec{x}))(f(\vec{x^\prime} - m(\vec{x^\prime})] \, .
$$
The mean function is usually the physical model included in addition to the GP model (i.e., $m_i^{\rm d}$). For a pure GP model, the mean function can be considered as zero. We need, then, to specify only the covariance function $k(\vec{x}, \vec{x^\prime})$, which needs to produce a valid covariance (i.e., a positive semi-definite symmetric) matrix.

\begin{figure}[t]
\begin{center}
\includegraphics[scale=1]{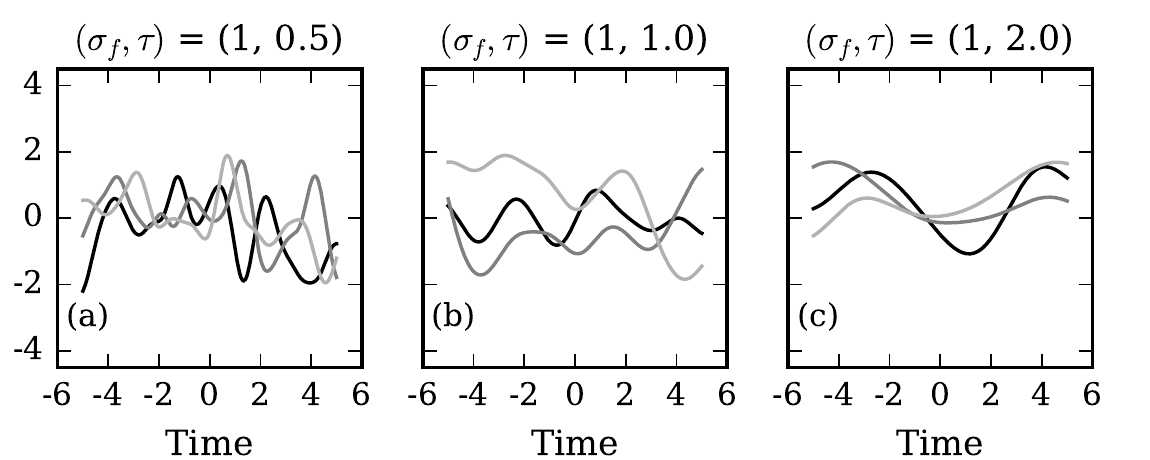}
\caption{Functions drawn from a GP with a squared-exponential kernel. The length scale parameter changes from 0.5 (a) to 1.0 (b) to 2.0 (c). As the length scale increases, the typical functions generated by the GP vary more slowly. In other words, the distance between two points in time that differ significantly increases with length scale.}
\label{fig.gpse}
\end{center}
\end{figure}

One very popular example of covariance matrix is the squared-exponential:
\begin{equation}
k(t_i, t_j)  = \sigma_f^2 \exp\left[-\frac{(t_i - t_j)^2}{2 \tau^2}\right]\,.
\label{eq.se}
\end{equation}
For simplicity we assumed time is the only input variable and changed the notation accordingly.  The function has two free parameters, $\sigma_f$ and $\tau$, called hyperparameters. We can sample from the GP defined by the covariance function in Eq.~(\ref{eq.se}), and without including yet any data point. We call this the \emph{prior} GP. The obtained sample is shown in Fig.~\ref{fig.gpse} for three different values of the length scale parameter $\tau$. In fact, to draw the functions in Fig.~\ref{fig.gpse} we have evaluated the covariance at a finite time vector $\vec{t}_*$ of length $n_*$, known as the input vector. The corresponding matrix of covariance values is denoted $K(t_*, t_*)$, where for simplicity we left out the vector notation for the input vector. 

The functions drawn in Fig.~\ref{fig.gpse} are samples from the prior GP. Usually, we are interested in knowing how these functions change when we condition the GP to some data.  For example, let us assume we have obtained a number of observations at times $t_i$, for $i = {1, 2, \ldots, n}$. At these points we know the values of the function, $f_i$, with some noise, which we assume normally distributed with variance $\sigma^2_i$, for $i = {1, 2, \ldots, n}$, i.e., the observation at time $t_i$ is $y_i = f_i + \epsilon_i$, where $\epsilon_i$ is distributed as $N(0, \sigma_i)$.  This data set is called the training set. We want to incorporate the new information and see how our process changes. \citet{rasmussenwilliams2005} explain that one way to do this is to produce a large number of samples from the prior GP and reject those that do not agree with the data. Of course, this is very inefficient computationally. Fortunately, the properties of GPs permit giving analytical formulae for the mean and covariance function of the posterior GP, $f_*$, conditioned to the data \citep[see eqs.~2.22--2.24 from][]{rasmussenwilliams2005}:
\begin{align}
E[f_*|\vec{t}, \vec{y}] &= K(X_*, X)\left[ K(X, X) + \sigma_n^2 I \right]^{-1}\vec{y}\,,\\
cov(f_*) &= K(X_*, X_*) - K(X_*, X)\left[K(X, X) + \sigma_n^2 I\right]^{-1} K(X, X_*)\,,
\end{align}
where the symbol $K(X, X_*)$ represents the $n \times x_*$ matrix of covariances evaluated at all pairs of training and input points. 

\begin{figure}[t]
\begin{center}
\includegraphics[scale=1]{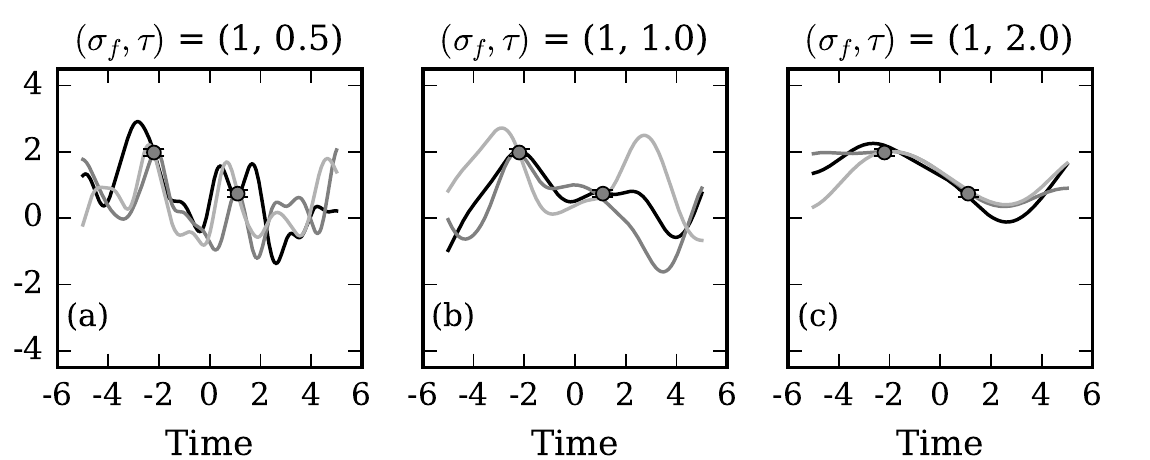}
\caption{Three draws from a GP obtained by conditioning the processes from Fig.~\ref{fig.gpse} with two training points, with time coordinates, $t_1=-2.2$, $t_2=1.1$, and observed values $y_1=1.98$, $y_2=0.74$. The variance of the noise term $\epsilon$ is 0.01 for both points. As the distance between the point becomes comparable with the decay time scale, the variance of the GP becomes smaller between the points. In other words, the information provided by one point }
\label{fig.gpsepost}
\end{center}
\end{figure}

Figure~\ref{fig.gpsepost} shows three draws from the posterior GP conditioned to two observations, separated by 3.3 ``time units". When the decay time is much smaller than this separation (panel (a)), the GP has a large variance between the observations, while its variance is much smaller when the separation of the training points is comparable with the decay time (panel (c)). In other words, if the covariance function falls off rapidly, having observed at one point in time gives little information about the rest of the points. Note that independently of the chosen hyperparameter values, the posterior GPs always manage to produce functions that accurately reproduce the training points. This attests the flexibility of the GPs, and warns us about their ability to potentially absorb coherent signals such as the signature of planets.

\newcommand{\period}{\mathcal{P}}
\newcommand{\structure}{\eta}
Another covariance function widely used to model the effect of stellar activity is the quasi-periodic kernel,
\begin{equation}
k(t_i, t_j) = \sigma_f^2 \exp\left[-\frac{(t_i -t_j)^2}{2\tau^2} -\frac{2\sin^2\left(\pi (t_i-t_j)/\period\right)}{\structure^2}\right]\,,
\end{equation}
with four hyperparameters: the covariance amplitude $\sigma_f^2$, the decay timescale $\tau$, the pseudo-period $\period$ and the structure parameter $\structure$. We present samples from this process in Fig.~\ref{fig.gpqp}. 

\begin{figure}[t]
\begin{center}
\includegraphics[scale=1]{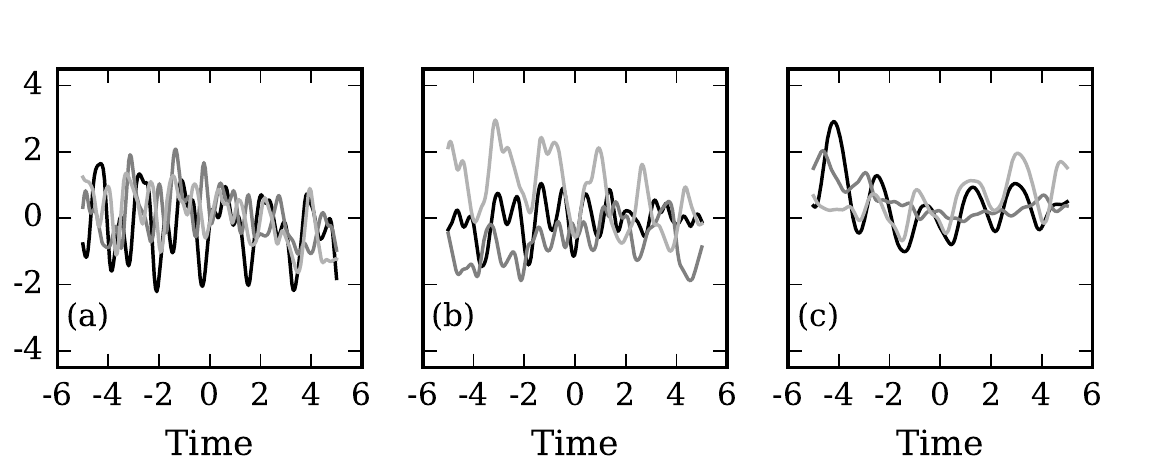}
\caption{Functions drawn from a GP with a pseudoperiodic kernel function, with amplitude $\sigma_f^2=1$, decay time scale $\tau=3$, pseudo period $\period=1.75$. The structure parameter is $\structure=0.75$ (a), $\structure=1.0$ (b), and $\structure=2.0$ (c). We can see that as the structure (hyper)parameter increases, the functions resemble a sinusoidal function more closely, and can more accurately describe a light curve dominated by a single long-lived spot. Functions with smaller structure parameter have a larger number of contributions and resemble more closely stars with more than one active region. For example, in panel (b) functions clearly have two periodic components that evolve independently.}
\label{fig.gpqp}
\end{center}
\end{figure}

The quasi-periodic kernel produces functions which resemble remarkably the light curve of spotted stars, and has been used to model both light curves and RV variations of active stars \citep[e.g.,][]{haywood2014, rajpaul2015}. The pseudo-period, or recurrence timescale, $\period$, can be identified with the stellar rotation period, which is often known from the light curve and in some situations from the radial-velocity time series. The decay time is associated with the time of the evolution of the active regions, which is harder to measure, but has been calibrated for solar-type stars using \textit{Kepler} photometry. Additionally, as shown in Fig.~\ref{fig.gpqp}, the structure parameter can be tuned to describe stars with a varying number of active regions. This kernel function allows therefore to describe a large variety of activity signals, making it very useful.

%\begin{svgraybox}
%CODE CODE CODE
%\end{svgraybox}

\section{Bayesian inference}
\label{sect.bayesian}
\newcommand{\like}{\mathcal{L}(\theta)}

This chapter on data modelling would not be complete without a description, however brief and summarised, of the techniques and methods used to obtain information on the model parameters from the data. The models presented so far are relatively simple and can in most cases be expressed by closed analytical formulae. However, the non-linear dependence on the model parameters and strong covariances between them makes the task of statistical inference challenging. In this section, we present a brief overview of the concepts involved in Bayesian statistical inference and provide a necessarily short introduction to a family of widely used algorithms: Markov chain Monte Carlo.

Statistical inference is the process by which we recover quantities of interest from available data, which are noisy or otherwise uncertain \citep[see, e.g.,][for an introduction addressed at astrophysicists and astronomers]{gregory, trotta2017}. The notion of probability is central in statistical inference. Statistics, sometimes called ``inverse probability", is focussed on learning about the underlying probabilistic processes that give rise to the observed data. This is done by learning about the numerical values of the model parameters. Unlike the classical (frequentist) concept of probability, which is related to the number of times an event occurs, and is therefore defined only for random variables, the Bayesian theory allows us to define the probability of any proposition. Indeed, the Bayesian concept of probability is related to the degree of belief in a given proposition, with the extremes being the concepts of True or False used in classical Aristotelian logic. Actually, Bayesian probability theory can be seen as an extension of logic allowing us to reason in the face of uncertainty. A wonderful description of the concepts underlying Bayesian probability theory and the process leading to considering it an extension of logic is given by  \citet{jaynes2003}. Many other advantages of the Bayesian approach have been identified and are usually quoted \citep[see, e.g.,][]{trotta2017}.

In Bayesian inference the prior knowledge of any proposition --- such as ``the orbital period of this planet is between 3.5 and 3.51 days'', or ``this star is orbited by six planetary companions'' --- is encoded in the prior probability distribution, $p(\theta | I)$, where $\theta$ represents a proposition, which can represent a given hypothesis or be related to the value of the model parameters for a given hypothesis. In any case, $I$ represents the knowledge available before taking into consideration the data. In what follows, we assume for simplicity that $\theta$ represents the parameter vector of a given model.

Once the data are incorporated in the analysis, the prior distribution transforms into the posterior distribution, $p(\theta | D, I)$. All the information provided by the data is encoded in this distribution. In general, the knowledge on $\theta$ changes when (new) data are incorporated. The way in which this change takes place is described by Bayes' theorem:
\begin{equation}
p(\theta | D, I) = \frac{p(D | \theta, I)p(\theta | I)}{p(D | I)}\,,
\end{equation}
where $D$ represents the data. The function $p(D | \theta, I) \equiv \like$ is called the likelihood function and describes how probable the current data are given a certain value of the model parameters, $\theta$. In the denominator, we find the marginal likelihood, $p(D | I)$, which is a constant term with respect to $\theta$ and ensures the correct normalisation of the posterior distribution. For the purpose of parameter inference, this can be safely ignored in most cases. On the other hand, the marginal likelihood has a central role in model comparison problems, which we do not cover here. The posterior distribution is therefore proportional  to the likelihood times the prior:
$$
p(\theta | D, I) \propto p(D | \theta, I)\,p(\theta | I)\,.
$$

\subsection{Likelihood function}
The likelihood function has therefore an important role in the inference process. A detailed description on how to construct likelihood functions is presented by \citet[][sect.~4.8]{gregory}. In particular, readers are referred to this book for a description on how to construct likelihoods with stochastic models.

Here we describe the simple case where the model is deterministic and the data errors are uncorrelated. Under these assumptions, one can deduce from Eq.~(\ref{eq.model}) that the probability distribution of a given datum, $d_i$, is equal to that of the error term\footnote{The derivation is not as trivial as one may think from considering Eq.~(\ref{eq.model}).} $e_i$. Proposition $D$ is simply the logical conjunction of propositions concerning individual data points: $D = d_1, d_2, \cdots, d_N$. Therefore, for independent errors, we have:
\begin{align}
\like = p(D | \theta, I) &= p(d_1, d_2, \cdots, d_N | \theta, I) \nonumber \\ &= p(e_1, e_2, \cdots, e_N | \theta, I) \nonumber \\ 
&=\prod_{i=1}^Np(e_i | \theta, I)\,.
\end{align}

In the case of normally distributed errors (Eq.~\ref{eq.normalerror}), the likelihood function has the form:
\begin{equation}
\like = \prod_{i=1}^N\frac{1}{\sqrt{2\pi}\sigma_i}\,\exp{-\frac{\left(d_i - m(t_i, \theta)\right)^2}{2\sigma_i^2}}\,,
\end{equation}
where $m(t_i, \theta)$ is the model function, where we have explicitly indicated its dependence on the independent variable (time), $t_i$, and the model parameter vector, $\theta$. Because of the large dynamical range of this likelihood function, it is usual to use its natural logarithm instead:
\begin{equation}
\log\like = -\frac{N}{2}\log(2\pi) - \frac{1}{2} \sum_{i=1}^N \log\sigma_i^2 - \frac{1}{2} \sum_{i=1}^N{\frac{\left(d_i - m(t_i, \theta)\right)^2}{\sigma_i^2}}\,.
\label{eq.loglike}
\end{equation}
The first term on the right-hand side of Eq.~(\ref{eq.loglike}) is a constant and is usually irrelevant for inference problems. The last term is proportional to the usual $\chi^2$ statistics. Note that if no model parameters appear in the data uncertainties, $\sigma_i$, maximising the likelihood function is equivalent to minimising the $\chi^2$ statistics. However, if data errors risk being underestimated, one might wish to include an additional (nuisance) parameter in the model to account for this (see Eq.~\ref{eq.normalerror2}). In that case the second term on the right-hand side acts as a counterbalance impeding the likelihood from becoming artificially large by increasing the size of the error bars.

When the uncertainties are correlated, or when a GP is included as part of the model, the log-likelihood function is:
\begin{equation}
\log\like = -\frac{N}{2}\log(2\pi) - \frac{1}{2} \log|K| - \frac{1}{2} \vec{r}^T\cdot K \cdot \vec{r}\,,
\label{eq.loglike}
\end{equation}
where $K$ is the covariance matrix, $|K|$ is its determinant and $\vec{r}$ is the vector of residuals obtained by subtracting the model prediction from the observed data points.

\subsection{Priors}
The prior distribution of $\theta$ is the second element needed to compute the posterior distribution. Prior distributions (or simply ``priors'') are a controversial feature of Bayesian statistics. While some see in them a way to naturally include relevant prior information in the analysis others deem prior distributions as an inherently subjective element of the theory. In any case, defining the priors of a given problem is usually difficult and tricky. Priors need to include all relevant information on a problem, which can range from the results of a previous analysis to the educated guess of a renown scientific figure. As priors need to have the form of probability distributions, a non-trivial problem of codification arises. While there is no universally accepted method to assign priors that convey information, there is a vast body of literature on ways to define ignorance priors.

%\paragraph{Ignorance priors}
Ignorance priors are intended to convey the least possible information about $\theta$. One could argue that absolute ignorance is an abstract concept and that some degree of information on a given parameter value is always available. To this \citet{jaynes2003} replies that ignorance priors play the role of the ``zero" in Bayesian statistics, i.e., no information, and as such have an important theoretical role. In more practical terms, ignorance priors are usually handy when one needs to specify the priors for parameters for which little can be said before considering the data.

The literature contains a large number of digressions on how ignorance priors should be defined. We give a brief description of the most popular of them here:

\begin{enumerate}
\item {\it Principle of indifference.} \citet{laplace1812} introduced a principle for assigning probabilities to a set of $n$ discrete hypotheses ${H_1, H_2, \ldots, H_n}$ when nothing allows to prefer one over the others. If the set is complete (i.e., they cover the entire set of possible outcomes), we have $p(H_i | I) = 1/n$, for all $i$. Although this method seems intuitive, it is not easily extended to continuous variables, besides lacking invariance to transformation.

\item {\it Invariance rules.} For a certain type of parameters, ``total ignorance" can be represented as invariance under certain transformations \citep{gregory}. For example, when the parameter represents a location, as for example the position $X$ of the largest tree along a river, total ignorance requires that the problem be invariant under a translation: $X^\prime = X + c$, for any constant $c$. This leads to the uniform prior:
\begin{equation}
p(X | I) = \text{constant}\,.
\end{equation}
For scale parameters, such as the Poisson rate of an unknown source or the lifetime of new bacteria, ignorance is reflected by a lack of information on the order of magnitude of the parameter (the bacteria can live from a few minutes to a few days or weeks). This can be translated to invariance under rescaling, $X^\prime = \alpha X$, which leads to a log-flat prior:
\begin{equation}
p(X | I) = \frac{\text{constant}}{X}\,.
\end{equation}
Although these rules are useful, not all parameters can be classified in any of these two categories. Besides, the invariance under transformations gives us the functional form of the prior distribution, but not its boundaries (normalisation). One needs still to define the extrema of the distributions, if proper priors are needed.

\item {\it Jeffreys-rule prior.} The rule advocated by \citet{jeffreys1946,jeffreys1961} uses the expected Fisher information for a given problem:
\begin{equation}
I(\theta) = E_D\left[-\frac{\partial^2}{\partial\theta_i\partial\theta_j}\log\like\right]\,,
\end{equation}
where the expectation is over the data\footnote{Real data $D$ are not available at the moment of specifying priors for the model parameters.}. Jeffreys's rule defines the prior for $\theta$ as
\begin{equation}
p(\theta | I) = \sqrt{I(\theta)}\,.
\end{equation}
Jeffreys's rule produces the optimal prior for one-dimensional models. Besides, it reproduces the results from the invariance rules. For example, the prior for the mean value of a normal model with known variance is a flat prior, and that of the scale of a normal with known mean value is a log-flat prior. However, Jeffreys-rule priors run into serious issues when applied to multi-dimensional parameters \citep{berger2015}, reducing drastically their applicability.
\end{enumerate}

Other methods involve the principle of maximum entropy \citep{jaynes1957, jaynes1968}. More recent developments include reference priors \citep{berger2009, berger2015}, where the Kullback--Leibler divergence between prior and posterior is maximised seeking a prior that maximises the expected difference between posterior and prior distributions. In one-dimensional problems, reference priors reduce to Jeffreys-rule priors, but their properties in multi-dimensional problems are better.

In any case, sensitivity analyses are warranted and should be performed systematically. However, as expressed by \citet[][sect.~3.3]{trotta2017}:
\begin{quote}
A sensitivity analysis should always be performed, i.e., change the prior in a
reasonable way and assess how robust the ensuing posterior is. Unfortunately, this is seldom done in the astrophysics and cosmology literature.
\end{quote}

\subsection{Sampling the posterior: Markov chain Monte Carlo}
Once the prior distribution is specified and the likelihood function is constructed, the posterior probability distribution (or simply the ``posterior'') can be obtained. In most practical cases it is not possible to perform inference on the posterior distribution analytically. For this reason, a series of algorithms to explore the posterior exist.  Chief among them, Markov chain Monte Carlo (MCMC) is a family of algorithms that can be employed to produce a sample of arbitrary size from the posterior. With a sample at hand, one can make inferences on the model parameters, define credible intervals, explore the marginalised distributions etc. An introduction to MCMC can be found in \citet{gregory}. Another useful reading is the appendix of \citet{tegmark2004}. A vast body of literature and online resources exist on the subject and are promptly available.

The basic idea behind MCMC is to construct a Markov chain starting at a given point in parameter space. New links are added to the chain following some stochastic algorithm. Under quite general conditions it can be shown that after some steps, the links of the chain are samples from the posterior distribution \citep[e.g.,][]{gregory}.  MCMC algorithms differ in the way the new chain links are produced. Many algorithms use a random walk process to generate a candidate link which is later accepted as part of the chain or not following some criterion. 

The most popular Random Walk MCMC algorithm in exoplanetary science is the Metropolis--Hastings (MH) algorithm \citep{metropolis1953, hastings1970}. In this algorithm, new steps for the chain are proposed by means of a random walk, usually the addition of a multivariate normal random variable to the current parameter vector, $\theta$, to produce a candidate link, $\theta^\prime$. The proposed new state, $\theta^\prime$, is then accepted with probability
$$
r = \min\left(1, \frac{p(\theta^\prime | D, I)}{p(\theta | D, I)}\,\frac{q(\theta| \theta^\prime)}{q(\theta^\prime | \theta)}\right)\,,
$$
where $q(y | x)$ is the probability of proposing state $y$ from state $x$. In the case of the normal proposal distribution, the ratio $q(\theta, \theta^\prime)/q(\theta^\prime, \theta)$ cancels out. In other words, if at the proposed step the posterior density increases with respect to $\theta$, it is accepted automatically. Otherwise, the algorithm accepts $\theta^\prime$ with probability $p(\theta^\prime | D, I)/p(\theta | D, I) < 1$.

The efficiency of the MH algorithm is often given by the choice of the proposal distribution $q(\cdot | \cdot)$. A na\"ive choice as the one described above usually makes the algorithm very inefficient for sampling correlated parameter space. More sophisticated adaptive algorithms are described in the literature \citep[e.g.,][]{haario2001, goodmanweare2010, diaz2014}. Another well-known issue of this MCMC algorithm is its inefficiency for sampling from multi-modal distributions. Readers are advised to consult the vast literature on this subject before developing their own MCMC code or using those available online.

\begin{acknowledgement}
The author thanks the organisers of the IVth Azores International Advanced School in Space Sciences and acknowledges the participants --- both lecturers and students --- for the quality of their work. The preparation of this lecture was carried out within the frame of the Swiss National Centre for Competence in Research ``PlanetS'' funded by the Swiss National Science Foundation (SNSF). The author acknowledges support by the Argentinian National Council for Research and Technology (CONICET).
\end{acknowledgement}

\bibliographystyle{apj}
\bibliography{biblio}

\begin{thebibliography}{}
\expandafter\ifx\csname natexlab\endcsname\relax\def\natexlab#1{#1}\fi

\bibitem[{{Agol} {et~al.}(2005){Agol}, {Steffen}, {Sari}, \&
  {Clarkson}}]{agol2005}
{Agol}, E., {Steffen}, J., {Sari}, R., \& {Clarkson}, W. 2005, \mnras, 359, 567

\bibitem[{Almenara {et~al.}(2016)Almenara, D{\'{\i}}az, Bonfils, \&
  Udry}]{almenara2016}
Almenara, J.~M., D{\'{\i}}az, R.~F., Bonfils, X., \& Udry, S. 2016, \aap, 595,
  L5

\bibitem[{{Almenara} {et~al.}(2015){Almenara}, {D{\'{\i}}az}, {Mardling},
  {Barros}, {Damiani}, {Bruno}, {Bonfils}, \& {Deleuil}}]{almenara2015}
{Almenara}, J.~M., {D{\'{\i}}az}, R.~F., {Mardling}, R., {et~al.} 2015, \mnras,
  453, 2644

\bibitem[{{Ballard} {et~al.}(2011){Ballard}, {Fabrycky}, {Fressin},
  {Charbonneau}, {Desert}, {Torres}, {Marcy}, {Burke}, {Isaacson}, {Henze},
  {Steffen}, {Ciardi}, {Howell}, {Cochran}, {Endl}, {Bryson}, {Rowe}, {Holman},
  {Lissauer}, {Jenkins}, {Still}, {Ford}, {Christiansen}, {Middour}, {Haas},
  {Li}, {Hall}, {McCauliff}, {Batalha}, {Koch}, \& {Borucki}}]{ballard2011}
{Ballard}, S., {Fabrycky}, D., {Fressin}, F., {et~al.} 2011, \apj, 743, 200

\bibitem[{{Barros} {et~al.}(2014){Barros}, {D{\'{\i}}az}, {Santerne}, {Bruno},
  {Deleuil}, {Almenara}, {Bonomo}, {Bouchy}, {Damiani}, {H{\'e}brard},
  {Montagnier}, \& {Moutou}}]{barros2014}
{Barros}, S.~C.~C., {D{\'{\i}}az}, R.~F., {Santerne}, A., {et~al.} 2014, \aap,
  561, L1

\bibitem[{Berger {et~al.}(2009)Berger, Bernardo, \& Sun}]{berger2009}
Berger, J.~O., Bernardo, J.~M., \& Sun, D. 2009, The Annals of Statistics, 905

\bibitem[{Berger {et~al.}(2015)Berger, Bernardo, Sun, {et~al.}}]{berger2015}
Berger, J.~O., Bernardo, J.~M., Sun, D., {et~al.} 2015, Bayesian Analysis, 10,
  189

\bibitem[{{Boisse} {et~al.}(2012){Boisse}, {Bonfils}, \& {Santos}}]{boisse2012}
{Boisse}, I., {Bonfils}, X., \& {Santos}, N.~C. 2012, \aap, 545, A109

\bibitem[{{Carter} {et~al.}(2011){Carter}, {Fabrycky}, {Ragozzine}, {Holman},
  {Quinn}, {Latham}, {Buchhave}, {Van Cleve}, {Cochran}, {Cote}, {Endl},
  {Ford}, {Haas}, {Jenkins}, {Koch}, {Li}, {Lissauer}, {MacQueen}, {Middour},
  {Orosz}, {Rowe}, {Steffen}, \& {Welsh}}]{carter2011}
{Carter}, J.~A., {Fabrycky}, D.~C., {Ragozzine}, D., {et~al.} 2011, Science,
  331, 562

\bibitem[{{Claret}(2000)}]{claret2000}
{Claret}, A. 2000, \aap, 363, 1081

\bibitem[{Correia {et~al.}(2010)Correia, Couetdic, Laskar, Bonfils, Mayor,
  Bertaux, Bouchy, Delfosse, Forveille, Lovis, Pepe, Perrier, Queloz, \&
  Udry}]{correia2010}
Correia, A. C.~M., Couetdic, J., Laskar, J., {et~al.} 2010, \aap, 511, A21

\bibitem[{{D{\'{\i}}az} {et~al.}(2014){D{\'{\i}}az}, {Almenara}, {Santerne},
  {Moutou}, {Lethuillier}, \& {Deleuil}}]{diaz2014}
{D{\'{\i}}az}, R.~F., {Almenara}, J.~M., {Santerne}, A., {et~al.} 2014, \mnras,
  441, 983

\bibitem[{{Doyle} {et~al.}(2011){Doyle}, {Carter}, {Fabrycky}, {Slawson},
  {Howell}, {Winn}, {Orosz}, {Pr\v{s}a}, {Welsh}, {Quinn}, {Latham}, {Torres},
  {Buchhave}, {Marcy}, {Fortney}, {Shporer}, {Ford}, {Lissauer}, {Ragozzine},
  {Rucker}, {Batalha}, {Jenkins}, {Borucki}, {Koch}, {Middour}, {Hall},
  {McCauliff}, {Fanelli}, {Quintana}, {Holman}, {Caldwell}, {Still},
  {Stefanik}, {Brown}, {Esquerdo}, {Tang}, {Furesz}, {Geary}, {Berlind},
  {Calkins}, {Short}, {Steffen}, {Sasselov}, {Dunham}, {Cochran}, {Boss},
  {Haas}, {Buzasi}, \& {Fischer}}]{doyle2011}
{Doyle}, L.~R., {Carter}, J.~A., {Fabrycky}, D.~C., {et~al.} 2011, Science,
  333, 1602

\bibitem[{{Dumusque} {et~al.}(2014){Dumusque}, {Boisse}, \&
  {Santos}}]{dumusque2014}
{Dumusque}, X., {Boisse}, I., \& {Santos}, N.~C. 2014, \apj, 796, 132

\bibitem[{Goldstein(1980)}]{goldstein}
Goldstein, H. 1980, Classical Mechanics. 2nd edition. (Addison-Wesley
  Publishing Company)

\bibitem[{Goodman \& Weare(2010)}]{goodmanweare2010}
Goodman, J., \& Weare, J. 2010, Communications in applied mathematics and
  computational science, 5, 65

\bibitem[{{Gregory}(2005)}]{gregory}
{Gregory}, P.~C. 2005, {Bayesian Logical Data Analysis for the Physical
  Sciences: A Comparative Approach with `Mathematica' Support} (Cambridge
  University Press, Cambridge, UK, 2005)

\bibitem[{Haario {et~al.}(2001)Haario, Saksman, \& Tamminen}]{haario2001}
Haario, H., Saksman, E., \& Tamminen, J. 2001, Bernoulli, 223

\bibitem[{Hastings(1970)}]{hastings1970}
Hastings, W. 1970, Biometrika, 57, 97

\bibitem[{{Haywood} {et~al.}(2014){Haywood}, {Collier Cameron}, {Queloz},
  {Barros}, {Deleuil}, {Fares}, {Gillon}, {Lanza}, {Lovis}, {Moutou}, {Pepe},
  {Pollacco}, {Santerne}, {S{\'e}gransan}, \& {Unruh}}]{haywood2014}
{Haywood}, R.~D., {Collier Cameron}, A., {Queloz}, D., {et~al.} 2014, \mnras,
  443, 2517

\bibitem[{{Holman} \& {Murray}(2005)}]{holmanmurray2005}
{Holman}, M.~J., \& {Murray}, N.~W. 2005, Science, 307, 1288

\bibitem[{{Holman} {et~al.}(2010){Holman}, {Fabrycky}, {Ragozzine}, {Ford},
  {Steffen}, {Welsh}, {Lissauer}, {Latham}, {Marcy}, {Walkowicz}, {Batalha},
  {Jenkins}, {Rowe}, {Cochran}, {Fressin}, {Torres}, {Buchhave}, {Sasselov},
  {Borucki}, {Koch}, {Basri}, {Brown}, {Caldwell}, {Charbonneau}, {Dunham},
  {Gautier}, {Geary}, {Gilliland}, {Haas}, {Howell}, {Ciardi}, {Endl},
  {Fischer}, {F{\"u}r{\'e}sz}, {Hartman}, {Isaacson}, {Johnson}, {MacQueen},
  {Moorhead}, {Morehead}, \& {Orosz}}]{holman2010}
{Holman}, M.~J., {Fabrycky}, D.~C., {Ragozzine}, D., {et~al.} 2010, Science,
  330, 51

\bibitem[{{Hut}(1981)}]{hut1981}
{Hut}, P. 1981, \aap, 99, 126

\bibitem[{Jaynes(1957)}]{jaynes1957}
Jaynes, E.~T. 1957, Physical review, 106, 620

\bibitem[{Jaynes(1968)}]{jaynes1968}
---. 1968, IEEE Transactions on systems science and cybernetics, 4, 227

\bibitem[{Jaynes(2003)}]{jaynes2003}
---. 2003, Probability Theory: The Logic of Science (Cambridge University
  Press)

\bibitem[{Jeffreys(1946)}]{jeffreys1946}
Jeffreys, H. 1946, in Proceedings of the Royal Society of London a:
  mathematical, physical and engineering sciences, Vol. 186, The Royal Society,
  453--461

\bibitem[{Jeffreys(1961)}]{jeffreys1961}
Jeffreys, H. 1961, Theory of probability, 3rd edn. (Oxford: Clarendon Press)

\bibitem[{{Jontof-Hutter} {et~al.}(2015){Jontof-Hutter}, {Rowe}, {Lissauer},
  {Fabrycky}, \& {Ford}}]{jontof-hutter2015}
{Jontof-Hutter}, D., {Rowe}, J.~F., {Lissauer}, J.~J., {Fabrycky}, D.~C., \&
  {Ford}, E.~B. 2015, \nat, 522, 321

\bibitem[{{Kipping}(2008)}]{kipping2008}
{Kipping}, D.~M. 2008, \mnras, 389, 1383

\bibitem[{{Kipping}(2010)}]{kipping2010}
---. 2010, \mnras, 408, 1758

\bibitem[{{Kipping}(2012)}]{kipping2012b}
---. 2012, \mnras, 427, 2487

\bibitem[{Laplace(1812)}]{laplace1812}
Laplace, P. 1812, Theorie Analytique des Probabilities. (Courcier)

\bibitem[{{Mandel} \& {Agol}(2002)}]{mandelagol2002}
{Mandel}, K., \& {Agol}, E. 2002, \apjl, 580, L171

\bibitem[{{Metropolis} {et~al.}(1953){Metropolis}, {Rosenbluth}, {Rosenbluth},
  {Teller}, \& {Teller}}]{metropolis1953}
{Metropolis}, N., {Rosenbluth}, A.~W., {Rosenbluth}, M.~N., {Teller}, A.~H., \&
  {Teller}, E. 1953, Journal of Chemical Physics, 21, 1087

\bibitem[{{Murray} \& {Correia}(2010)}]{murraycorreia2010}
{Murray}, C.~D., \& {Correia}, A.~C.~M. 2010, in Exoplanets, ed. S.~{Seager}
  (University of Arizona Press), 15--23

\bibitem[{{Murray} \& {Dermott}(2000)}]{murraydermott2000}
{Murray}, C.~D., \& {Dermott}, S.~F. 2000, {Solar System Dynamics} ({Cambridge
  University Press})

\bibitem[{{Nesvorn{\'y}} {et~al.}(2013){Nesvorn{\'y}}, {Kipping}, {Terrell},
  {Hartman}, {Bakos}, \& {Buchhave}}]{nesvorny2013}
{Nesvorn{\'y}}, D., {Kipping}, D., {Terrell}, D., {et~al.} 2013, \apj, 777, 3

\bibitem[{{Rajpaul} {et~al.}(2015){Rajpaul}, {Aigrain}, {Osborne}, {Reece}, \&
  {Roberts}}]{rajpaul2015}
{Rajpaul}, V., {Aigrain}, S., {Osborne}, M.~A., {Reece}, S., \& {Roberts}, S.
  2015, \mnras, 452, 2269

\bibitem[{Rasmussen \& Williams(2005)}]{rasmussenwilliams2005}
Rasmussen, C.~E., \& Williams, C. K.~I. 2005, Gaussian Processes for Machine
  Learning (Adaptive Computation and Machine Learning) (The MIT Press)

\bibitem[{{Seager} \& {Mall{\'e}n-Ornelas}(2003)}]{seagermallen-ornelas2003}
{Seager}, S., \& {Mall{\'e}n-Ornelas}, G. 2003, \apj, 585, 1038

\bibitem[{Tegmark {et~al.}(2004)Tegmark, Strauss, Blanton, Abazajian, Dodelson,
  Sandvik, Wang, Weinberg, Zehavi, Bahcall, Hoyle, Schlegel, Scoccimarro,
  Vogeley, Berlind, Budavari, Connolly, Eisenstein, Finkbeiner, Frieman, Gunn,
  Hui, Jain, Johnston, Kent, Lin, \& Nakajima}]{tegmark2004}
Tegmark, M., Strauss, M.~A., Blanton, M.~R., {et~al.} 2004, Phys. Rev. D, 69,
  103501

\bibitem[{Trotta(2017)}]{trotta2017}
Trotta, R. 2017, ArXiv e-prints, arXiv:1701.01467

\bibitem[{{Winn}(2010)}]{winn2008}
{Winn}, J.~N. 2010, in Exoplanets, ed. S.~{Seager} (University of Arizona
  Press), 55--77

\bibitem[{{Zahn}(1977)}]{zahn77}
{Zahn}, J.-P. 1977, \aap, 57, 383

\end{thebibliography}

\end{document}